\documentstyle[psfig,preprint,prl,aps]{revtex}
\def\half{\frac{1}{2}}
\def\vev{v.e.v.}
\def\gsim{~{\rlap{\lower 3.5pt\hbox{$\mathchar\sim$}}\raise 1pt\hbox{$>$}}\,}
\def\lsim{~{\rlap{\lower 3.5pt\hbox{$\mathchar\sim$}}\raise 1pt\hbox{$<$}}\,}
\def\a#1{\alpha_#1}
\def\ap{\alpha_1'}
\def\lh{\lambda_H}
\def\lk{\lambda_K}
\def\ln{\lambda_N}
\def\mg{m_{3/2}}
\def\lam#1{\lambda_#1}
\def\lamp{\lambda'_1}
\def\m#1{\tilde m_#1}
\def\mp{\tilde m_1'}
\def\MHs#1{M_{H_#1}^2}
\def\MSs{M_S^2}
\def\QQ{Q_Q}
\def\QU{Q_{U^c}}
\def\QD{Q_{D^c}}
\def\QL{Q_{L}}
\def\QN{Q_{N^c}}
\def\QE{Q_{E^c}}
\def\QH#1{Q_{H_#1}}
\def\QS{Q_S}
\def\QK{Q_K}
\def\QKc{Q_{K^c}}
\def\Up{\rm U'(1)}
\def\ASSQ{{\rm [SU(3)]^2U'(1)}}
\def\AWWQ{{\rm [SU(2)]^2U'(1)}}
\def\AYYQ{{\rm [U(1)]^2U'(1)}}
\def\AYQQ{{\rm U(1)[U'(1)]^2}}
\def\AQQQ{{\rm [U'(1)]^3}}

\def\PRD#1#2#3{Phys. Rev. {\bf D#1}, #3 (19#2)}
\def\NPB#1#2#3{Nucl. Phys. {\bf B#1}, #3 (19#2)}
\def\PTP#1#2#3{Prog. Theor. Phys. {\bf #1}, #3 (19#2)}

\def\PLB#1#2#3{Phys. Lett. {\bf B#1}, #3 (19#2)}
\def\PRL#1#2#3{Phys. Rev. Lett. {\bf #1}, #3 (19#2)}
\def\PRep#1#2#3{Phys. Rep. {\bf #1}, #3 (19#2)}
\begin{document}
\draft
\preprint{OCHA-PP-149} 

\title{ 
Supersymmetric Extension of the Standard Model  \\ 
with Naturally Stable Proton
}
\author{
Mayumi Aoki
\footnote{Address after April 1:  Theory Group, KEK, Tsukuba, 
Ibaraki 305-0801, Japan.}
}
\address{
 Graduate School of Humanities and Sciences, Ochanomizu University  \\
Otsuka 2-1-1, Bunkyo-ku, Tokyo 112-8610, Japan  \\
}
\author{
Noriyuki Oshimo
}
\address{
 Department of Physics, Ochanomizu University  \\
Otsuka 2-1-1, Bunkyo-ku, Tokyo 112-8610, Japan  \\
}
\date{\today}
\maketitle
\begin{abstract}

      A new supersymmetric standard model based on $N=1$ supergravity 
is constructed, aiming at natural explanation for the proton stability 
without invoking an ad hoc discrete symmetry through $R$ parity.  
The proton is protected from decay by an extra U(1) gauge symmetry.   
Particle contents are necessarily increased to be free from anomalies, 
making it possible to incorporate the superfields for 
right-handed neutrinos and an SU(2)-singlet Higgs boson.  
The vacuum expectation value of this Higgs boson, which induces  
spontaneous breakdown of the U(1) symmetry, yields   
large Majorana masses for the right-handed neutrinos, 
leading to small masses for the ordinary neutrinos.
The linear coupling of SU(2)-doublet Higgs superfields, 
which is indispensable to the superpotential of the minimal supersymmetric 
standard model, is replaced by a trilinear coupling of the 
Higgs superfields, so that there is no mass parameter in the superpotential.   
The energy dependencies of the model parameters are studied, 
showing that gauge symmetry breaking is induced by radiative corrections.  
Certain ranges of the parameter values compatible with phenomena at the 
electroweak energy scale can be derived from universal values of 
masses-squared and trilinear coupling constants 
for scalar fields at a very high energy scale.  

\end{abstract}
\pacs{12.60.Cn, 11.30.Fs, 12.60.Jv}

\narrowtext

\section{Introduction}

     The standard model (SM) well describes particle physics 
below the electroweak energy scale.  
However, various theoretical considerations 
suggest that some extension 
of the SM be necessary for physics above that energy scale.  
Various models therefore have been proposed, 
some of which being studied extensively.  
Among them extensions with supersymmetry \cite{nilles} 
are considered most plausible.   
In particular, the minimal supersymmetric standard model (MSSM) is  
usually treated as their standard theory around the electroweak energy scale.  

     The MSSM inherits most of the successful features of the SM, 
while the extension being minimal.  
However, this model suffers one serious setback, which 
has been often passed over.  
In the SM, the proton is protected from decay naturally by gauge symmetry.  
On the other hand,  in the MSSM, the gauge symmetry allow the 
interactions of dimension four which do not conserve 
baryon and/or lepton numbers.  
Unless there exists some reason to forbid these 
interactions, the proton decays in an unacceptably short time.  
Therefore, a discrete symmetry is usually imposed 
on the MSSM through $R$ parity, which is however merely 
an ad hoc symmetry.  

     A convincing reason for the proton stability could be provided 
by an extra gauge symmetry.  
Although such a symmetry around the electroweak energy scale  
is subjected to many phenomenological constraints, 
they still show room to allow a U(1) gauge symmetry.   
Several supersymmetric models with an extra U(1) symmetry 
therefore have been discussed \cite{weinberg,hall,oshimo,font,chamseddine}, 
aiming at natural explanation for a long lifetime of the proton.  
However, these models are accompanied by some arbitrariness 
in construction, which might reduce reliability of the reasoning 
for the proton stability.  

     In this paper, the supersymmetric extension 
of the SM with an extra U(1) gauge symmetry is studied  
within the framework of a model coupled to $N=1$ supergravity.  
In addition to the proton lifetime, the MSSM involves 
problems on the neutrino masses and the linear coupling of Higgs 
superfields noted later.   
Requiring a model to solve these problems consistently in a minimal 
extension, its particle contents and superpotential are determined 
rather uniquely \cite{aoki}.  
In sizable ranges of the model parameters, the scalar potential 
appropriately gives a vacuum of SU(3)$\times$U$_{EM}$(1) gauge symmetry 
below the electroweak energy scale. 
Phenomenological predictions of the model are compatible 
with experimental results.  
A typical mass scale of scalar particles is of order 1 TeV, 
which can account for the smallness of the electric 
dipole moments (EDMs) of the neutron and the electron, another 
problem of the MSSM.   

     The energy dependencies of the model parameters are also discussed  
by analyzing renormalization group equations (RGEs).  
Taking the masses-squared of scalar fields all positive at a high 
energy scale, those for some Higgs fields become sufficiently 
small at lower energy scales to induce  
the breakdowns of the extra U(1) and electroweak gauge symmetries.  
In the model coupled to $N=1$ supergravity 
the masses-squared and the trilinear coupling 
constants for scalar fields are considered to respectively have 
universal values at a very high energy scale.  
This scenario can be realized in this model.  

     In constructing the model, we take into account 
the problems of the neutrino masses and the Higgs linear coupling 
as well as that of the proton stability.  
The former is raised by non-vanishing masses of the neutrinos 
suggested from experiments for atmospheric and solar neutrinos, 
such as at the Super-Kamiokande \cite{kamiokande}.  
The MSSM or the SM can have Yukawa couplings for neutrino Dirac masses,  
if right-handed neutrinos are naively included.    
However, these fields are inert for the transformations of the 
gauge groups and their existence is not prescribed by the model.  
Furthermore, the extreme lightness of the neutrinos  
may require some explanation.  
Although this lightness could be attributed to large Majorana masses of 
the right-handed neutrinos, their origin is not clarified.  

     The latter problem is posed by a mass parameter of 
the $\mu$ term, a linear coupling of 
the Higgs superfields in the superpotential of the MSSM \cite{kim}, 
which is indispensable for correct breaking of electroweak gauge symmetry.    
This mass parameter $\mu$ 
should have a magnitude of order the electroweak energy scale.  
The other mass parameters in the model are traced back to 
supersymmetry-soft-breaking terms of the Lagrangian and thus related to 
the gravitino mass which may be of order the electroweak energy scale.  
On the other hand, the $\mu$ parameter is contained in the 
supersymmetric term and its magnitude may be given arbitrarily.  
It is natural that there is no mass parameter 
in the superpotential and the role of the $\mu$ term is assumed 
by another effective $\mu$ term.    

     These problems could also be solved by introducing an extra U(1) 
gauge symmetry.  
Imposing a new gauge symmetry yields chiral 
and trace anomalies within the particle contents of the MSSM.  
For canceling the anomalies, new superfields are necessarily incorporated,  
among which those for right-handed neutrinos and an SU(2)-singlet 
Higgs boson may be included.   
The extra U(1) symmetry could be broken above the 
electroweak energy scale by a vacuum expectation 
value (\vev) of this Higgs boson, which may provide a source of 
Majorana masses for the right-handed neutrinos and 
the effective $\mu$ parameter.  

     This paper is organized as follows.  
In Sect. 2 we construct a model, in which the proton is 
adequately stable, the ordinary neutrinos have non-vanishing but 
small masses, and the effective $\mu$ term is contained.  
In Sect. 3 the vacuum structure of the model is discussed, 
paying particular attention to experimental constraints 
on an extra neutral gauge boson.  
In Sect. 4 the behavior of the model parameters for different energy scales 
is analyzed through the RGEs to examine the radiative breaking  
of gauge symmetry and the supersymmetry breaking by $N=1$ supergravity.    
Conclusions and discussions are given in Sect. 5.  

\section{Model}

     Particle contents of the model are constrained by 
the requirements of proton stability, neutrino masses, 
and an effective $\mu$ term.   
We also keep the extension of the SM as minimal as possible.  
The neutrino masses necessitate superfields for 
right-handed neutrinos, which are denoted by $N^c$.  
For the effective $\mu$ term an SU(3)$\times$SU(2)$\times$U(1) singlet 
superfield $S$ is included.   
In addition, new colored superfields $K$ and $K^c$ are necessary 
for canceling a chiral anomaly, as shown later.  
In Table \ref{particles} we list the left-handed chiral superfields 
contained in the model with their quantum numbers under SU(3), 
SU(2), U(1), and $\Up$ gauge transformations.  
The extra U(1) gauge symmetry is  
denoted by $\Up$, for which the charges of superfields 
are expressed as $\QQ$, $\QU$, etc..  
In order not to yield chiral and trace anomalies within 
the standard gauge symmetries, we assign opposite U(1) charges 
$Y_K$ and $-Y_K$ to $K$ and $K^c$.   
The index $i$ $(i=1,2,3)$ stands for the generation, 
while the indices $j$ of $H_1$ and $H_2$, $k$ of $S$, 
and $l$ of $K$ and $K^c$ are for possible 
multiplication to be determined by cancellation of the anomalies.  

     The superpotential should contain the couplings 
$H_1QD^c$, $H_2QU^c$, $H_1LE^c$, and $H_2LN^c$ to generate   
masses for quarks and leptons.  
The $\mu$ term can be replaced by the coupling $SH_1H_2$, 
provided that the scalar component of $S$ has a non-vanishing \vev.  
The Dirac masses of the neutrinos may be comparable to 
those of the charged leptons, unless the Yukawa coupling constants 
are extremely small.   
However, the ordinary neutrino masses are suppressed 
by giving large Majorana masses to the right-handed neutrinos, 
which can be accomplished, without another new field, 
by including the coupling $SN^cN^c$.  
These couplings provide constraints on the $\Up$ charges 
of the superfields:  
\begin{eqnarray}
    \QH1+\QQ+\QD &=& 0,   
\label{HQD}  \\
    \QH2+\QQ+\QU &=& 0,   
\label{HQU}  \\
    \QH1+\QL+\QE &=& 0,   
\label{HQN}  \\
    \QH2+\QL+\QN &=& 0,   
\label{HQE}  \\
    \QS+\QH1+\QH2 &=& 0,   
\label{SHH}  \\
    \QS+\QN+\QN &=& 0.   
\label{SNN}  
\end{eqnarray}
If colored superfields are only those which correspond to 
the quarks of the SM, the $\ASSQ$ anomaly-free condition 
with Eqs. (\ref{HQD}) and (\ref{HQU}) 
gives the relation $\QH1+\QH2=0$. 
The linear coupling $H_1H_2$ are not forbidden in the superpotential,  
and thus the model inevitably has a mass parameter of unknown origin.   
Therefore, new colored superfields should be included 
to solve the problem of the $\mu$ term.   
Although there are various candidates for such superfields, 
according to the 'minimal' postulate, we incorporate a pair 
of superfields in the fundamental representations of the SU(3) 
group, $K$ and $K^c$.  
Then, their fermion components should have large masses 
from a phenomenological viewpoint, which is fulfilled 
by allowing the coupling $SKK^c$.   
This coupling leads to another constraint 
\begin{equation}
    \QS+\QK+\QKc = 0.    
\label{SKK}
\end{equation}
The $\ASSQ$ anomaly-free condition and Eqs. (\ref{HQD}), 
(\ref{HQU}), (\ref{SHH}), and (\ref{SKK}) fix  
the number $n_l$ of pairs for $K$ and $K^c$ at three, 
which agrees with the number of the generation for quarks and leptons.  

     The number $n_j$ of pairs for $H_1$ and $H_2$ and the 
number $n_k$ for $S$ are determined by the freedom from chiral anomalies 
for $\AWWQ$, $\AYYQ$ and a trace anomaly for U$'$(1) with 
Eqs. (\ref{HQD})-(\ref{SHH}), (\ref{SKK}).  
These constraints are satisfied by either of the three sets of numbers 
and U(1) charge:  
\begin{eqnarray} 
({\rm A})\ &&  n_j=4, n_k=5, Y_K=0,  \nonumber \\ 
({\rm B})\ &&  n_j=3, n_k=3, Y_K=\pm \frac{1}{3},  \nonumber \\ 
({\rm C})\ &&  n_j=2, n_k=1, Y_K=\pm \frac{\sqrt{2}}{3}.  \nonumber  
\end{eqnarray}
However, the solution (A) does not satisfy the $\AYQQ$ anomaly-free 
condition with Eq. (\ref{SNN}).  
The solution (C) gives irrational U(1) charges for $K$ and $K^c$.  
The solution (B) is free from the $\AYQQ$ anomaly, and 
also satisfies the remaining $\AQQQ$ anomaly-free condition.  
Therefore, a plausible solution is uniquely given by the set (B).  
The numbers $n_j$ and $n_k$ again become equal to 
the number of the generation.  
   
     A further constraint comes from the stable proton.  
The allowed value of $Y_K$ for the U(1) charges of $K$ and $K^c$ 
is now either 1/3 or $-1/3$.   
However, the proton stability by gauge symmetry is only 
achieved for $Y_K=1/3$ \cite{oshimo,font}.  
For $Y_K=-1/3$, the particle contents of one 
generation can be embedded in the fundamental representation of the E$_6$ group.   
Unless a discrete symmetry is imposed, the baryon and/or lepton 
numbers are violated by couplings of dimension four, 
such as $U^cD^cK^c$ and $LQK^c$,  
which induce an unacceptably fast decay of the proton.  
On the other hand, for $Y_K=1/3$, allowed couplings 
of dimension four are only those which have already been taken into 
account, i.e. $H_1QD^c$, $H_2QU^c$, $H_1LE^c$, $H_2LN^c$, 
$SH_1H_2$, $SN^cN^c$, and $SKK^c$.  
Baryon number is conserved while lepton number is not, which 
is sufficient for the proton stability.  
The lowest dimension couplings of baryon-number 
violation are given by the D terms of $QQU^{c*}E^{c*}$, 
$QQD^{c*}N^{c*}$, and $QU^{c*}D^{c*}L$, which are of dimension six.  

     Under all the anomaly-free conditions 
and Eqs. (\ref{HQD})-(\ref{SKK}), the $\Up$ charges of the 
superfields are expressed in terms of two independent variables.  
All the superfields are triplicated, and the anomalies are canceled 
in each generation.  
The generators $Y'$ and $Y$ of $\Up$ and U(1),  respectively, are 
required to be orthogonal, ${\rm Tr}[Y'Y]=0$.  
Then, the $\Up$ charges of the superfields are 
determined up to a normalization factor.  
For definiteness, hereafter, the $\Up$ charges are normalized to the 
U(1) charges as ${\rm Tr}[Y'^2]={\rm Tr}[Y^2]$, 
which are shown in Table \ref{charges}.  

     The superpotential which contains all the couplings consistent 
with gauge symmetry and renormalizability is given by
\begin{eqnarray}
W &=& \eta_d^{ijk} H_1^iQ^jD^{ck} + \eta_u^{ijk}H_2^iQ^jU^{ck} 
 + \eta_e^{ijk} H_1^iL^jE^{ck} + \eta_\nu^{ijk} H_2^iL^jN^{ck}   \nonumber \\
&& + \lambda_N^{ijk}S^iN^{cj}N^{ck} + \lambda_H^{ijk}S^iH_1^jH_2^k 
  + \lambda_K^{ijk}S^iK^jK^{ck}, 
\label{superpotential}
\end{eqnarray}
where $\eta_d$, $\eta_u$, $\eta_e$, $\eta_\nu$, $\ln$, $\lh$, and $\lk$ 
represent dimensionless constants.    
Contraction of group indices is understood.  
In the MSSM without the discrete symmetry through $R$ parity, 
the couplings $D^cD^cU^c$, $LQD^c$, $LLE^c$, $H_1H_1E^c$, and $LH_2$ 
are allowed, leading to non-conservation of baryon and lepton numbers.  
Here, these couplings are forbidden by the U$'$(1) gauge symmetry.  
The proton decay could only occur through the operators of 
dimension six, being suppressed at least by a huge mass to 
the second power.  
As long as this mass scale is not much smaller than the Planck mass,   
the proton becomes adequately stable.   
The couplings of the superpotential 
are all cubic, and there is no mass parameter. 

     We assume that supersymmetry is broken through the 
ordinary mechanism based on $N=1$ supergravity.  
Supergravity is spontaneously broken in a hidden sector at 
the Planck mass scale, and then supersymmetry in an observable sector
is broken softly.  
At lower energy scales, the Lagrangian of the observable 
sector consists of a supersymmetric part and 
a supersymmetry-soft-breaking part prescribed 
by gauge symmetry and superpotential.  
The soft-breaking part contains mass terms for 
gauge fermions, and trilinear couplings and mass terms for scalar bosons,    
\begin{eqnarray}
{\cal L_{SB}} &=& 
 -\half\left (\m3\bar\lam3\lam3 + \m2\bar\lam2\lam2 + 
  \m1\bar\lam1\lam1 + \mp\bar\lamp\lamp \right ) \nonumber \\ 
&& -\mg\left(A_d^{ijk}\eta_d^{ijk}H_1^iQ^jD^{ck} 
                 + A_u^{ijk}\eta_u^{ijk}H_2^iQ^jU^{ck}  
  + A_e^{ijk}\eta_e^{ijk} H_1^iL^jE^{ck} 
+ A_\nu^{ijk}\eta_\nu^{ijk} H_2^iL^jN^{ck}  \right. \nonumber \\
&& \left. + B_N^{ijk}\lambda_N^{ijk}S^iN^{cj}N^{ck} 
+ B_H^{ijk}\lambda_H^{ijk}S^iH_1^jH_2^k 
  + B_K^{ijk}\lambda_K^{ijk}S^iK^jK^{ck} \right) + {\rm H.c.} \nonumber \\
&& - M_{Q^i}^2 |Q^i|^2-M_{U^{ci}}^2 |U^{ci}|^2-M^2_{D^{ci}} |D^{ci}|^2 
- M_{L^i}^2|L^i|^2-M_{N^{ci}}^2|N^{ci}|^2-M^2_{E^{ci}}|E^{ci}|^2 \nonumber \\
&& -M_{H_1^i}^2|H_1^i|^2-M_{H_2^i}^2|H_2^i|^2-M_{S^i}^2|S^i|^2
- M_{K^i}^2|K^i|^2-M_{K^{ci}}^2|K^{ci}|^2.  
\label{softbreaking}
\end{eqnarray}
Here $\lam3$, $\lam2$, $\lam1$, and $\lamp$ represent  
gauge fermions for SU(3), SU(2), U(1), and U$'$(1), respectively. 
Scalar bosons are denoted by the same symbols as the corresponding 
superfields.  
With $\mg$ being the gravitino mass, the coefficients $A_d$, $A_u$, 
$A_e$, $A_\nu$, $B_N$, $B_H$, and $B_K$ are dimensionless.  
At high energy scales not much lower than the Planck mass, 
the masses-squared of scalar fields are all around $\mg^2$ and positive.  
The trilinear coupling constants for scalar fields are also  
approximately the same.  
Around the electroweak energy scale, some of these parameters 
differ significantly  
from the high-energy values through large quantum corrections.  

\section{Vacuum Structure}

     The Lagrangian of our model has 
SU(3)$\times$SU(2)$\times$U(1)$\times$U$'$(1) 
gauge symmetry, which must be spontaneously broken down to 
SU(3)$\times$U$_{EM}$(1) symmetry.  
This breaking could be achieved by \vev s for the scalar 
components of $H_1^i$, $H_2^i$, and $S^i$.  
We discuss the vacuum structure of the model by examining 
the scalar potential.   
Hereafter, we adopt the same notation for the superfields and 
their scalar components.  

     Although the scalar potential could contain all 
of $H_1^i$, $H_2^i$, and $S^i$ and thus its general analysis  
is complicated, it may be simplified under certain assumptions.    
If the couplings between different generations are not significant, 
$H_2^3Q^3U^{c3}$ of the third generation has a large coefficient 
related to the mass of the top quark.  
The mass-squared of $H_2^3$ then receives large negative 
contributions through quantum corrections and becomes small 
around the electroweak energy scale.   
As a result, $H_1^3$ and $H_2^3$ can have non-vanishing \vev s 
and assume the role of two Higgs doublets in the MSSM.   
For the first two generations, on the other hand, 
such couplings for $H_1^i$ or $H_2^i$ have small coefficients, 
so that the masses-squared of these scalar fields are kept around $\mg^2$.   
If the coefficient of $S^3K^3K^{c3}$ is large, the mass-squared of 
$S^3$ is also driven small.  
Although there is no phenomenological information about $S^iK^iK^{ci}$,  
a hierarchy of their coefficients could well exist.  
Among the three scalar fields $S^i$, 
one scalar field $S^3$ alone may have a non-vanishing \vev.  
We thus assume that only $H_1^3$, $H_2^3$, and $S^3$ 
can have non-vanishing \vev s.   
Quantum corrections to the masses-squared of other scalar 
fields are small, keeping them around $\mg^2$.  
The masses-squared of $K^i$, $K^{ci}$, and $N^{ci}$ 
receive non-negligible negative contributions from the D-term of U$'$(1) 
when the gauge symmetry is broken spontaneously.  
However,  
the positive contributions from the supersymmetry-soft-breaking terms 
in Eq. (\ref{softbreaking}) can dominate over and prevent 
these scalar fields from getting non-vanishing \vev s.  

     Assuming the above simplification, the scalar potential is given by 
\begin{eqnarray}
V &=& \frac{1}{8}g_2^2\left( |H_1|^2+|H_2|^2\right) ^2
     +\frac{1}{8}g_1^2\left( |H_1|^2-|H_2|^2\right) ^2 \nonumber  \\
    & & +\frac{1}{72}{g'_1}^2\left( 4|H_1|^2+|H_2|^2
                       -5|S|^2\right) ^2  \nonumber  \\
    & & -\left( \frac{1}{2}g_2^2-|\lh|^2\right) |H_1H_2|^2 
              +|\lh|^2\left( |H_1|^2+|H_2|^2\right)|S|^2  \nonumber \\
    & & +\left( B_H\lh\mg SH_1H_2+{\rm H.c.}\right)
                      +\MHs1|H_1|^2+\MHs2|H_2|^2+\MSs|S|^2,  
\label{potential}
\end{eqnarray}
where the generation indices are left out.   
With group indices being expressed, $H_1H_2$ is written as 
$\epsilon_{ab}H_{1a}H_{2b}$, so that holds an equation  
$|H_1H_2|^2=|H_1|^2|H_2|^2-|H_1^{\dagger}H_2|^2$ .  
The gauge coupling constants for SU(2), U(1), and U$'$(1) are denoted 
by $g_2$, $g_1$, and $g'_1$, respectively.   

     We now discuss the \vev s of the Higgs fields $\langle H_1\rangle$, 
$\langle H_2\rangle$, and $\langle S\rangle$.  
For any values of $\langle H_1\rangle$ and $\langle H_2\rangle$, 
the complex phase of $\langle S\rangle$ has a value which gives an equality 
$B_H\lh\mg\langle SH_1H_2\rangle =-|B_H\lh\mg\langle SH_1H_2\rangle |$.  
For given values of $\langle |H_1|^2\rangle$ and $\langle |H_2|^2\rangle$, 
the \vev\ $\langle |H_1H_2|^2\rangle$ becomes maximum at 
$\langle H_1^{\dagger}H_2\rangle =0$.  
Therefore, a condition $g_2^2>2|\lh|^2$ guarantees electric charge conservation.  
Differently from the MSSM, there is no direction for the \vev s where 
their quartic terms are absent in the scalar potential.  
The potential gives a stable vacuum irrespectively of the 
supersymmetry-soft-breaking terms.  
Redefining the global phases of the Higgs fields so as to 
give $B_H\lh=-|B_H\lh|$, the \vev s $v_1$, $v_2$, and $v_s$ of the neutral 
components of $H_1$, $H_2$, and $S$, respectively, 
may be taken real and non-negative.   
If these \vev s are all non-vanishing, extremum conditions are given by  
\begin{eqnarray}
&& \frac{1}{8}(g_2^2+g_1^2)(v_1^2-v_2^2)v_1
       +\frac{1}{18}g_1'^2(4v_1^2+v_2^2-5v_s^2)v_1  \nonumber  \\
&& +\half|\lh|^2(v_2^2+v_s^2)v_1-\frac{1}{\sqrt{2}}|B_H\lh\mg|v_2v_s 
       +\MHs1 v_1 = 0,  
\label{extremumH1}  \\
&&  -\frac{1}{8}(g_2^2+g_1^2)(v_1^2-v_2^2)v_2
       +\frac{1}{72}g_1'^2(4v_1^2+v_2^2-5v_s^2)v_2 \nonumber \\
&& +\half|\lh|^2(v_1^2+v_s^2)v_2-\frac{1}{\sqrt{2}}|B_H\lh\mg|v_1v_s
       +\MHs2 v_2 = 0,  
\label{extremumH2}  \\
&& -\frac{5}{72}g_1'^2(4v_1^2+v_2^2-5v_s^2)v_s
       +\half|\lh|^2(v_1^2+v_2^2)v_s \nonumber \\
&& -\frac{1}{\sqrt{2}}|B_H\lh\mg|v_1v_2+\MSs v_s = 0.
\label{extremumS}
\end{eqnarray}
It turns out that the solution of these simultaneous equations 
is unique, if exists.  
The true vacuum is either at such a point or at 
a point where at least one \vev\ vanishes, being determined by 
the potential energies of those points.  

     The \vev s of the Higgs bosons have to satisfy phenomenological 
constraints coming from experiments for the gauge bosons.  
The $W$-boson mass has been measured precisely.  
The $Z$ boson for SU(2)$\times$U(1) and the $Z'$ boson for U$'$(1) 
are mixed and their mass-squared matrix is given by  
\begin{eqnarray}
      & & \left(\matrix{ M_Z^2 & M_{ZZ'}^2 \cr 
            M_{ZZ'}^2 & M_{Z'}^2} \right), 
\label{matrix}  \\
M_Z^2 &=& \frac{1}{4}(g_2^2+g_1^2)(v_1^2+v_2^2), \\ 
M_{Z'}^2 &=& \frac{1}{36}g_1'^2(16v_1^2+ v_2^2+25v_s^2),    \\
M_{ZZ'}^2 &=& \frac{1}{12}g_1'\sqrt{g_2^2+g_1^2}(4v_1^2-v_2^2).    
\label{mzzp} 
\end{eqnarray}
Two massive neutral gauge bosons, which are denoted by 
$Z_1$ and $Z_2$ ($M_{Z_1}<M_{Z_2}$), are predicted.    
The measured mass for the $Z$ boson of the SM should be taken 
as the mass of $Z_1$.  
The experimental lower bound on the mass of a new neutral gauge boson 
is about 600 GeV \cite{cdf}.   
According to detailed analyses of various experiments for an extra 
gauge boson \cite{cho}, the mixing between $Z$ and $Z'$ is small.  
Defining a mixing parameter by $R=(M_{ZZ'}^2)^2/M_{Z}^2M_{Z'}^2$, 
a bound $R\lsim 10^{-3}$ is roughly obtained.    
The \vev s can also be constrained by the lightest 
Higgs boson mass, whose experimental bound is given by 
$M_{H^0}\gsim 80$ GeV \cite{higgs}.   
Since its predicted mass by the tree-level potential in Eq. (\ref{potential})   
could be altered to become larger by several tens of GeV 
through one-loop quantum corrections,  
we conservatively put a constraint $M_{H^0}>50$ GeV to the tree-level mass.   

\begin{figure}
\caption{The regions for $\MHs1$ and $\MHs2$ consistent with 
the experimental constraints.  The other parameters are taken for 
$|B_H\lh\mg|=0.1$ TeV and $|\lh|$=0.1, 0.3.   
}
\label{massrange}
\vspace {2 cm}
\begin{center}
\leavevmode 
\psfig{file=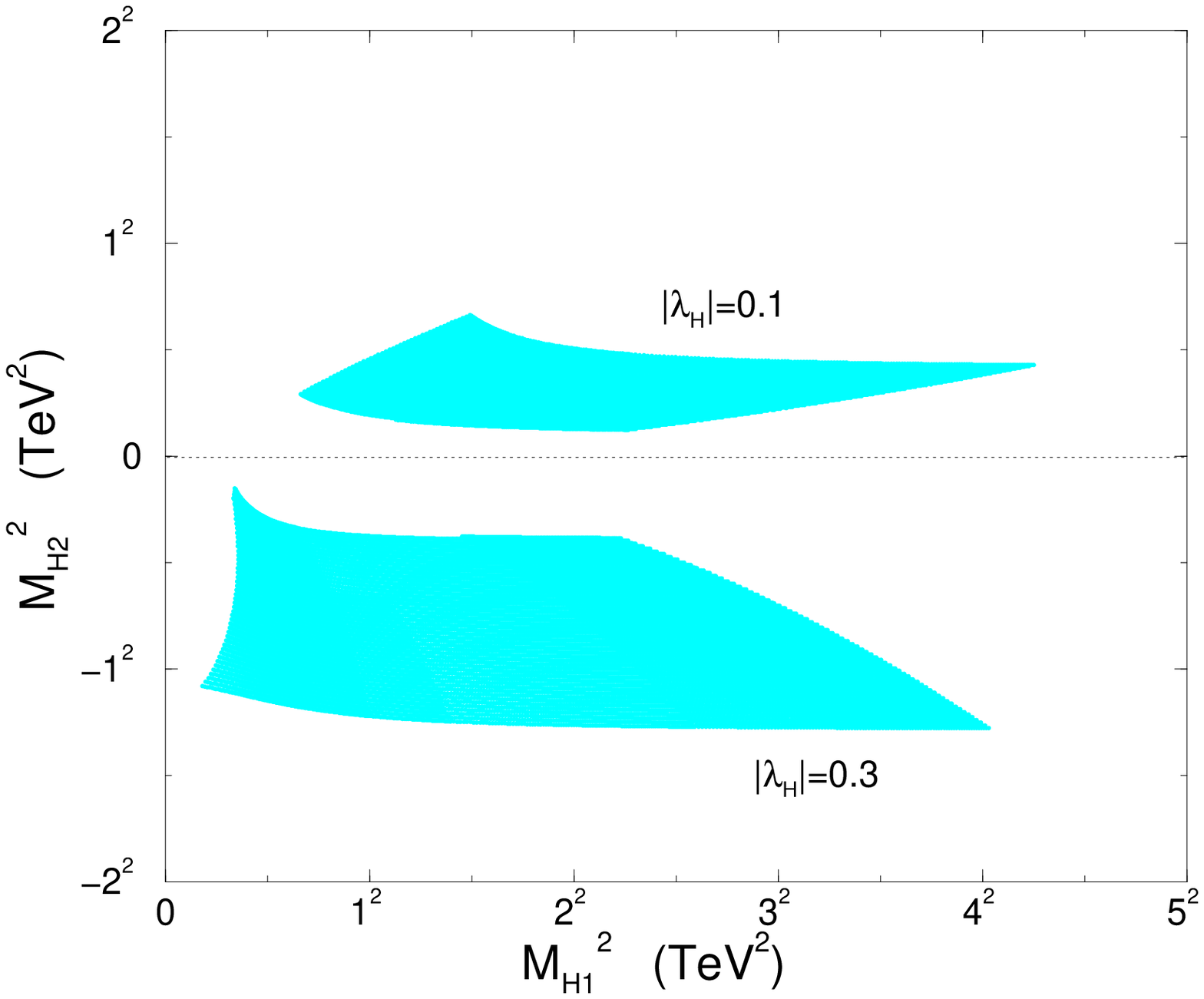,height=10cm}
\end{center}
\end{figure}

     The scalar potential is analyzed numerically.  
For independent coefficients of the potential we choose 
$|\lh|$, $|B_H\lh\mg|$, $\MHs1$, $\MHs2$, and $\MSs$.  
In Fig. \ref{massrange} we show 
the regions for $\MHs1$ and $\MHs2$ where the \vev s are 
compatible with the above constraints.  
We have also imposed the constraints 
$1\leq v_2/v_1\leq35$ and $M_{Z_2}\leq 2000$ GeV.  
With $|B_H\lh\mg|$ being 0.1 TeV,  
$|\lh|$ is set for 0.1 and 0.3, which correspond to the upper   
and lower regions, respectively, 
For given values of $\MHs1$, $\MHs2$, $|\lh|$, and $|B_H\lh\mg|$, 
the remaining parameter $\MSs$ is so determined as to make 
the $W$-boson mass coincident with the measured value.   
The gauge coupling constant for U$'$(1) is taken for $g_1'=g_1$.   
Owing to the constraints from $M_{Z_2}$ and $R$, in wide regions 
$\MHs1$ is larger than (1 TeV)$^2$.  
The value of $\MHs2$ is generally smaller than $\MHs1$ in magnitude.  
The region for $|\lh|=0.3$ with $\MHs1\lsim (500$ GeV)$^2$ 
corresponds to $v_2/v_1\lsim 2$.  
A rough estimate of Eq. (\ref{extremumH2}) shows that the sign of  
$\MHs2$ is positive for $|\lh|^2<(5/36)g_1'^2$ while 
$\MHs2$ has either sign for larger values of $|\lh|$.  
The value of $\MSs$ is smaller than $\MHs2$ and always negative.  
As $|B_H\lh\mg|$ increases, the allowed values for 
$\MHs1$ become larger, which is seen from Eq. (\ref{extremumH1}).  
If the upper limit for $M_{Z_2}$ is lifted, wider parameter regions become allowed.  
However, as the scale of the mass-squared parameters increases, 
more fine-tuning of the parameters becomes inevitable for electroweak 
symmetry breaking.  
For having the correct vacuum, 
large differences among $\MHs1$, $\MHs2$, and $\MSs$ are necessary, 
which could well occur under our assumption for supersymmetry breaking.   

     In Table \ref{masses} we present 
four examples for the values of $\MHs1$, $\MHs2$, and $\MSs$ 
in the allowed regions of Fig. \ref{massrange}.   Also shown are 
the resultant values for $v_2/v_1$, $v_s$, 
$M_{Z_2}$, $R$, and the masses of the physical Higgs bosons.  
These Higgs-boson masses have been calculated, assuming for definiteness
that the mass eigenstates are formed by the Higgs fields $H_1$, $H_2$, 
and $S$ without mixing with the other fields
of $H_1^i$, $H_2^i$, and $S^i$.
Therefore, there are three mass eigenstates for the neutral scalar 
bosons $H^0$, one for the neutral pseudoscalar boson $A^0$, and one 
for the charged scalar boson $H^\pm$.  
One neutral scalar boson is light, whereas the others have large masses.  
The mixing parameter $R$ vanishes for $v_2/v_1=2$, as seen from 
Eq. (\ref{mzzp}).  

\begin{figure}
\caption{The $\MSs$ dependency of the predicted $W$-boson mass for 
the examples (i), (ii), (iii), and (iv) in Table \ref{masses}.
$M_{S0}^2$ represents the values of $\MSs$ which give the measured 
mass of the $W$ boson $M_{Wexp}$.}
\label{finetuning}
\vspace {2 cm}
\begin{center}
\leavevmode 
\psfig{file=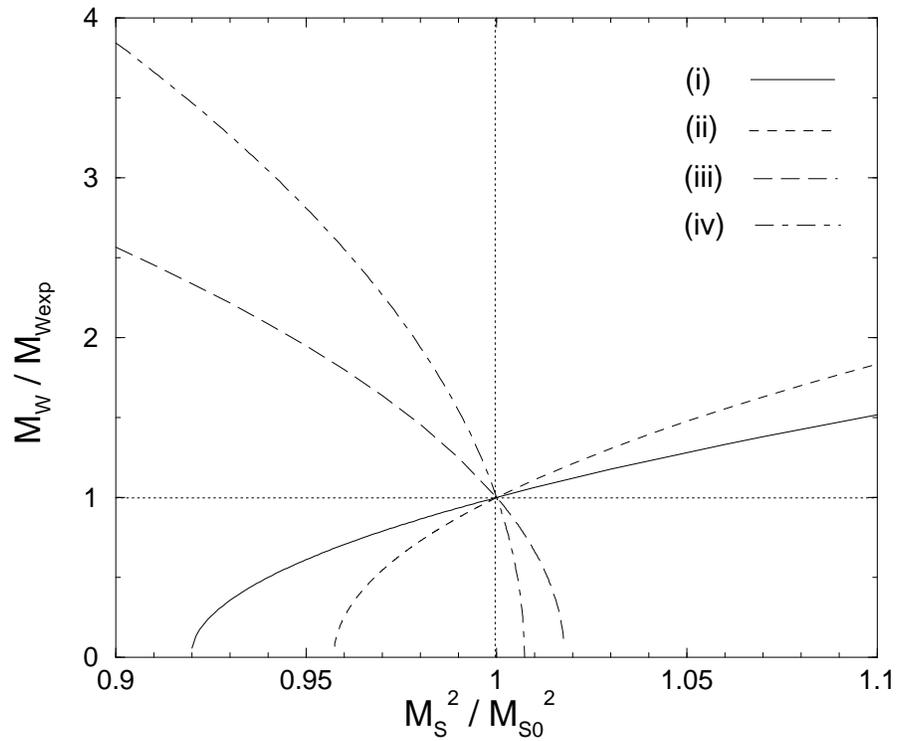,height=10cm}
\end{center}
\end{figure}

     The large mass difference between $Z_1$ and $Z_2$  
requires in some degree fine-tuning for the parameters in the potential.  
Since these two masses are different from each other by one 
order of magnitude, it is generally necessary to adjust the values of 
the mass-squared parameters $\MHs1$, $\MHs2$, $\MSs$ and 
the coupling constants $\lh$, $B_H$ within the accuracy of order $10^{-2}$.  
In Fig. \ref{finetuning}, for the four examples in Table \ref{masses},  
the ratio of a predicted 
$W$-boson mass to the experimental value is depicted 
as a function of $\MSs$ normalized to its proper 
value which yields the correct $W$-boson mass.   
If the value of one parameter alone 
in the potential is deviated by order of $10^{-1}$, 
the resultant v.e.v.s lead to a $W$-boson mass 
different from its experimental value by a factor or more.   

     The neutrinos have both Dirac and Majorana masses.  
Neglecting the generation mixing, the mass matrix for the 
left-handed and right-handed neutrinos becomes  
\begin{equation}
      \left(\matrix{ 0 & -\eta_\nu v_2/\sqrt{2} \cr 
        -\eta_\nu v_2/\sqrt{2} & \sqrt{2}\ln v_s} \right),   
\label{neutrino}
\end{equation}
whose lighter mass eigenvalue is approximately given by 
$m_{\nu 1}=|\eta_\nu|^2v_2^2/2\sqrt{2}|\ln|v_s$.   
With $|\ln|=0.2$ and $v_s=3$ TeV,  
$m_{\nu 1}$ becomes about 59 eV for $|\eta_\nu|v_2=10$ MeV and 
0.59 eV for $|\eta_\nu|v_2=1$ MeV.  
Even if the Yukawa coupling constants for the neutrino Dirac masses are  
of the same order as that for the electron, the observed ordinary 
neutrinos have tiny masses which could be a reason for the recent experimental 
results suggesting neutrino oscillations.   

     The coupling $SH_1H_2$ serves as the $\mu$ term and 
an effective $\mu$ parameter is given by $\mu=\lh v_s/\sqrt{2}$.   
For $|\lh|=0.2$ and $v_s=3$ TeV, $|\mu|$ is approximately 420 GeV.  
The parameter $\mu$ can have an appropriate magnitude for the 
electroweak symmetry breaking.   
This parameter also affects the masses of the charginos and the neutralinos.  
Assuming that these particles are formed by the fermion components 
of $H_1$, $H_2$, and $S$, as well as the SU(2), U(1), and U$'$(1) 
gauge fermions, there exist two mass eigenstates for the charginos 
and six mass eigenstates for the neutralinos.   
Provided that the gauge fermions for SU(2), U(1), and U$'$(1) 
receive masses of order 100 GeV from the supersymmetry-soft-breaking 
terms in Eq. (\ref{softbreaking}), 
the masses of the lighter chargino and the lightest neutralino  
become of order 100 GeV.  

     This model contains new particles which 
are not predicted by the MSSM.  
As already noted, the gauge-Higgs sector involves 
extra a neutral gauge boson, a neutral Higgs boson, and two neutralinos.  
For the lepton sector, there appear a heavy neutrino in each generation.  
Correspondingly the scalar neutrinos are duplicated.  
The interactions arising from the coupling $SN^cN^c$ do not conserve 
lepton number.  
In addition, the superfields $H_1^i$, $H_2^i$, $S^i$ with 
$i=1,2$ and $K^j$, $K^{cj}$ with $j=1,2,3$ are newly introduced.  
The masses of their fermion components are 
generated by the couplings to $H_1^3$, $H_2^3$, and $S^3$, 
and become of order $0.1-1$ TeV.  
The lightest fermion among $K^j$ and $K^{cj}$ is stable.   
As well as by collider experiments, such a stable particle may be 
explored by other methods to search for its relics in the 
universe, e.g. anomalous nuclei in sea water.  
However, these methods depend on the relic density, 
whose theoretical prediction is plagued by various 
uncertainties for non-perturbative effects, cosmology, and so on.  
Since the scalar components of $H_1^i$ and $H_2^i$ 
couple to quarks and leptons, non-trivial constraints 
are imposed on their coupling constants from the 
viewpoint of flavor-changing neutral current.   
However, these scalar particles are rather heavy, 
so that the constraints are not so stringent as usually thought.    
If the couplings $S^iH_1^3H_2^3$, $S^3H_1^iH_2^3$, or 
$S^3H_1^3H_2^i$ are not neglected, some or all of 
the scalar or fermion components for 
$H_1^i$, $H_2^i$, and $S^i$ are mixed with the Higgs bosons,  
charginos, or neutralinos, leading to an enlargement of the particles 
belonging to the gauge-Higgs sector.      

     The MSSM has another problem on the EDMs of the neutron 
and the electron, which can be explained in this model.  
If the squark and slepton masses are of order 
100 GeV and the $CP$-violating phases intrinsic in the model 
are not suppressed, these EDMs are predicted to be much larger 
than their experimental upper bounds.    
However, a typical scale of the squark and slepton masses 
in this model, which are considered not much different from the mass of $H_1$, 
is larger than 1 TeV.  
The EDMs then lie within the experimental bounds without fine-tuning 
the $CP$-violating phases to be very small \cite{kizukuri}.  
If these phases are not suppressed, the interactions of the charginos or 
the neutralinos generally induce sizable $CP$ violation in their 
production or decay processes.  

\section{Energy Dependence} 

     The parameter values of the model 
change according to the relevant energy scale.  
Analyzing their energy dependencies, 
we discuss whether gauge symmetry breaking is induced by 
radiative corrections.  
We also examine the scenario of universal values for  
the masses-squared and the trilinear coupling 
constants of scalar fields at a very high energy scale.  
For simplicity, the generation mixing of the particle fields are neglected.  

     The evolution of the parameters concerning the energy-scale change 
are described by RGEs, which are given in Appendix A. 
It is seen from those equations that $\MHs2$ increases
as the energy scale becomes high, 
owing to a large Yukawa coupling constant $\eta_u$ for the top quark.  
If the coupling constant $\lk$ is around unity, $\MSs$ also increases.  
Consequently, the mass-squared parameters can all have large 
positive values at high energy scales, even if they are small at a  
low energy scale as discussed in the previous section.   
The SU(2)$\times$U(1)$\times$U$'$(1) symmetry is spontaneously broken 
through radiative corrections.  
The experimental values of the gauge coupling constants suggest 
that these constants are not unified at the energy scale 
for possible grand unification.  
This gauge unification could be achieved by 
incorporating one additional pair of SU(2)-doublet chiral superfields.    
However, such a pair form a gauge-singlet linear coupling 
and thus ruin the model by necessitating a mass parameter of 
unknown origin.   
Although the particle contents are not embedded in the fundamental 
representation of the E$_6$ group, the masses and the coupling 
constants evolve similarly to those in the E$_6$ models. 
Some features of these models \cite{ma} apply to the 
present model, and vice versa.   

     We now  numerically examine the evolution of the parameters.  
Taking the masses-squared and the trilinear coupling constants 
of the scalar fields for common values $\mg^2$ and $A$ 
at a high energy scale $M_X$, we evaluate the parameters 
at a low energy scale $M$.  
For definiteness, we set $M_X$ for $10^{17}$ GeV and 
$M$ for $5\times 10^{2}$ GeV.  
Assuming an equality $g_1=g_1'$, all the gauge coupling constants 
are determined independently of the parameters.     
The masses of the gauge fermions are also determined, if their values 
are given at some energy scale.  
Since the gauge groups are not unified in our model, these masses at $M_X$ are 
generally different from each other.  
However, they are considered nevertheless to be of the same 
order of magnitude.  
We therefore put their values equal at $M_X$, 
$\m3=\m2=\m1=\mp\equiv\tilde m$, for simplicity.    
The Yukawa coupling constants $\eta_u$, $\eta_d$, $\eta_\nu$, 
$\eta_e$, $\lambda_N$, $\lambda_H$, and $\lambda_K$ at $M_X$, 
which are specified by attaching an index '$X$', 
are independent of each other.  

\begin{figure}
\caption{The low-energy values of $\eta_u$, $\ln$, $\lh$, $\lk$ 
for $\eta_u^X=0-1$ and $\ln^X=\lh^X=\lk^X=0.2$.}
\label{yukawa}
\vspace {2 cm}
\begin{center}
\leavevmode 
\psfig{file=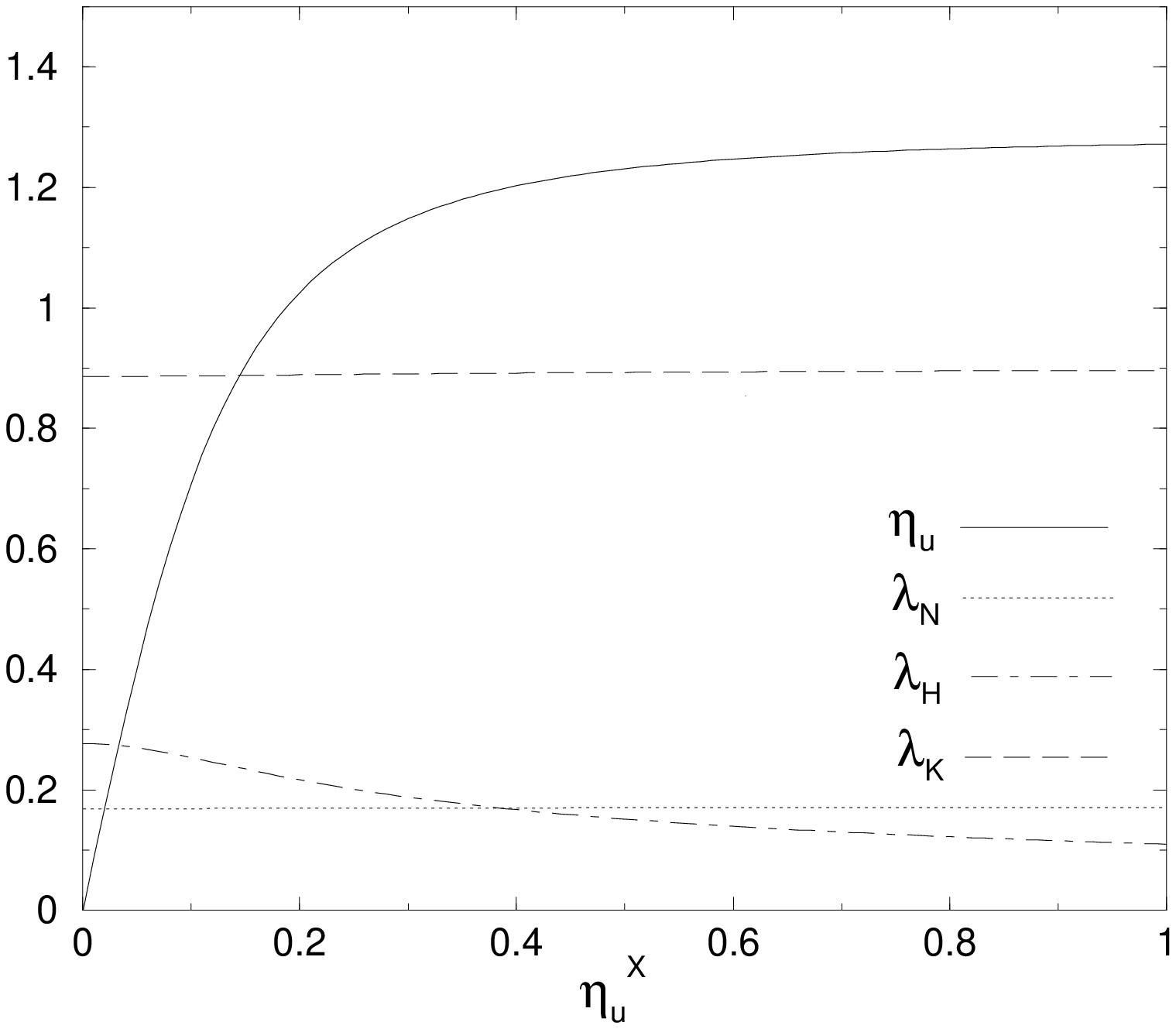,height=10cm}
\end{center}
\end{figure}

     In Fig. \ref{yukawa} the values of $\eta_u$, $\ln$, $\lh$, and $\lk$ 
at $M$ are depicted as functions of $\eta_u^X$.   
We have taken $\ln^X=\lh^X=\lk^X=0.2$ and 
$\eta_d^X=\eta_\nu^X=\eta_e^X=0$.  
The magnitude of $\eta_u$ and $\lambda_K$ become large at the low 
energy scale, while the energy dependencies of $\ln$ and $\lh$ are 
not significant.   
The evolution of each Yukawa coupling constant is not 
affected much by the other parameters.  
For generating an appropriate mass for the top quark, 
$|\eta_u^X|$ should be larger than 0.1.   
The condition $g_2^2>2|\lh|^2$ for electric charge conservation 
at the low energy scale is satisfied for $|\lh^X|\lsim 0.5$.  

\begin{figure}
\caption{The low-energy value of $B_H$ for the parameter sets 
(a), (b), and (c) in Table \ref{parameters}.  $\ln^X=\lh^X=0.2$.}
\label{trilinear}
\vspace {2 cm}
\begin{center}
\leavevmode 
\psfig{file=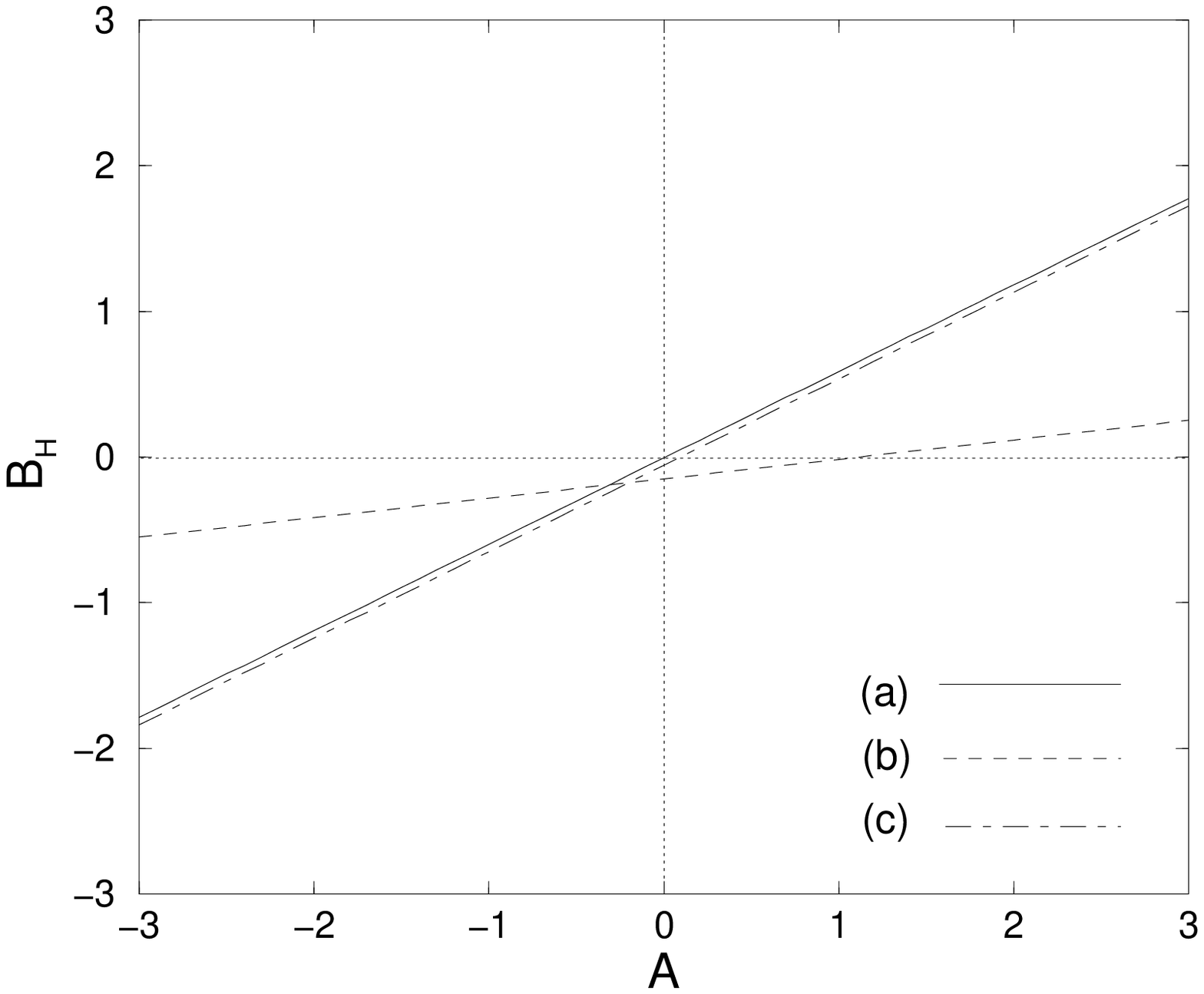,height=10cm}
\end{center}
\end{figure}

     In Fig. \ref{trilinear} we show the trilinear 
coupling constant $B_H$ as a function of $A$ for three 
sets of $\eta_u^X$, $\lk^X$, and $\tilde m/\mg$ given 
in Table \ref{parameters}.    
For the other non-vanishing input parameters, 
we take $\ln^X=\lh^X=0.2$.  
The magnitude of $A$ is constrained as  
$|A|< 3$ in order not to induce incorrect breaking of gauge symmetry.  
The value of $B_H$ is of order unity in wide ranges of $A$ except 
narrow ranges where $|B_H|$ is much smaller than unity.      
The $\tilde m$ dependence of $B_H$ is negligible.    

\begin{figure}
\caption{The low-energy values of $\MHs1$, $\MHs2$, and $\MSs$ 
for the parameter sets (a), (b), and (c) in Table \ref{parameters}.  
$\ln^X=\lh^X=0.2$,  $\mg=2$ TeV.}
\label{rgemass}
\vspace {2 cm}
\begin{center}
\leavevmode 
\psfig{file=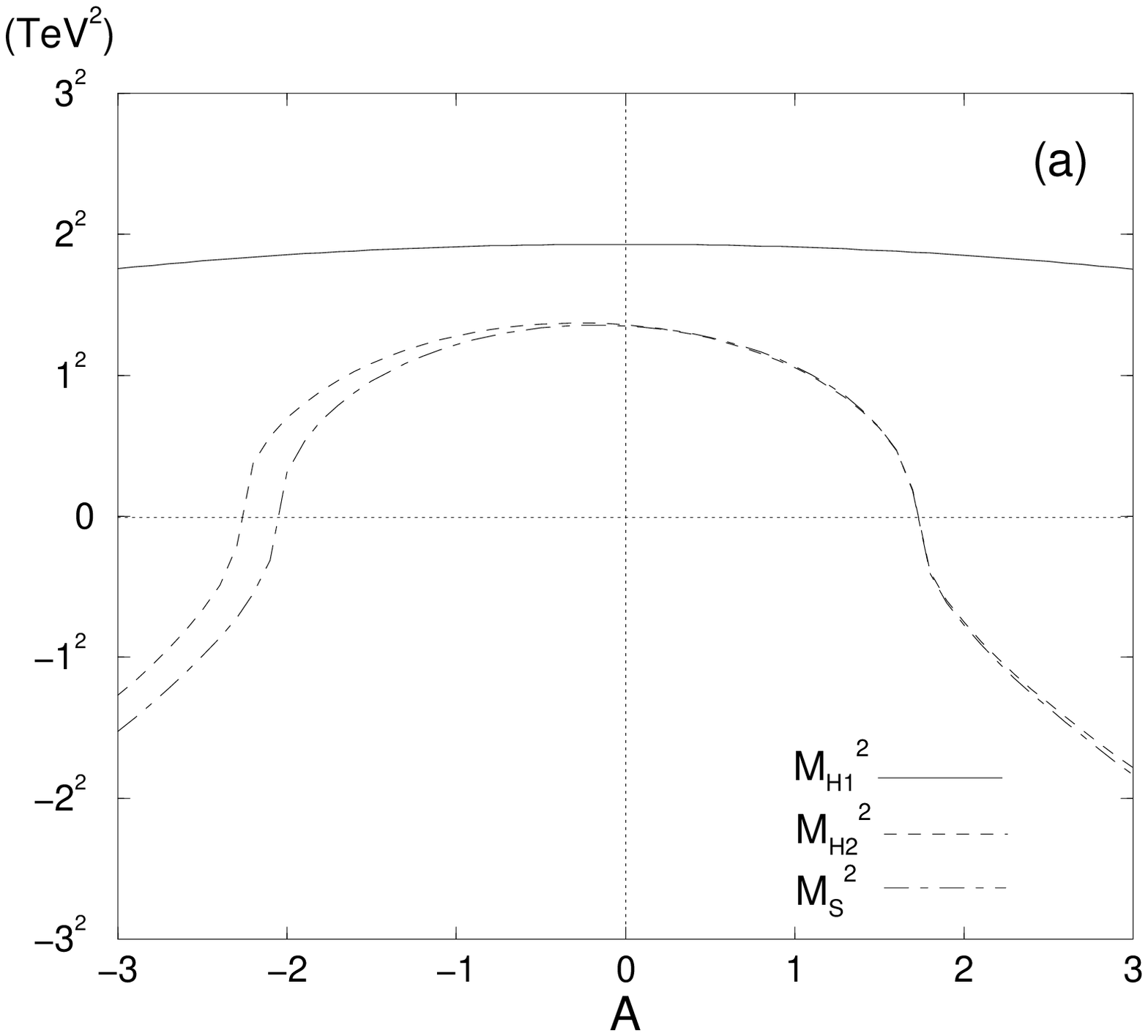,height=10cm}
\end{center}

\pagebreak 

\vspace {2 cm}
\begin{center}
\leavevmode 
\psfig{file=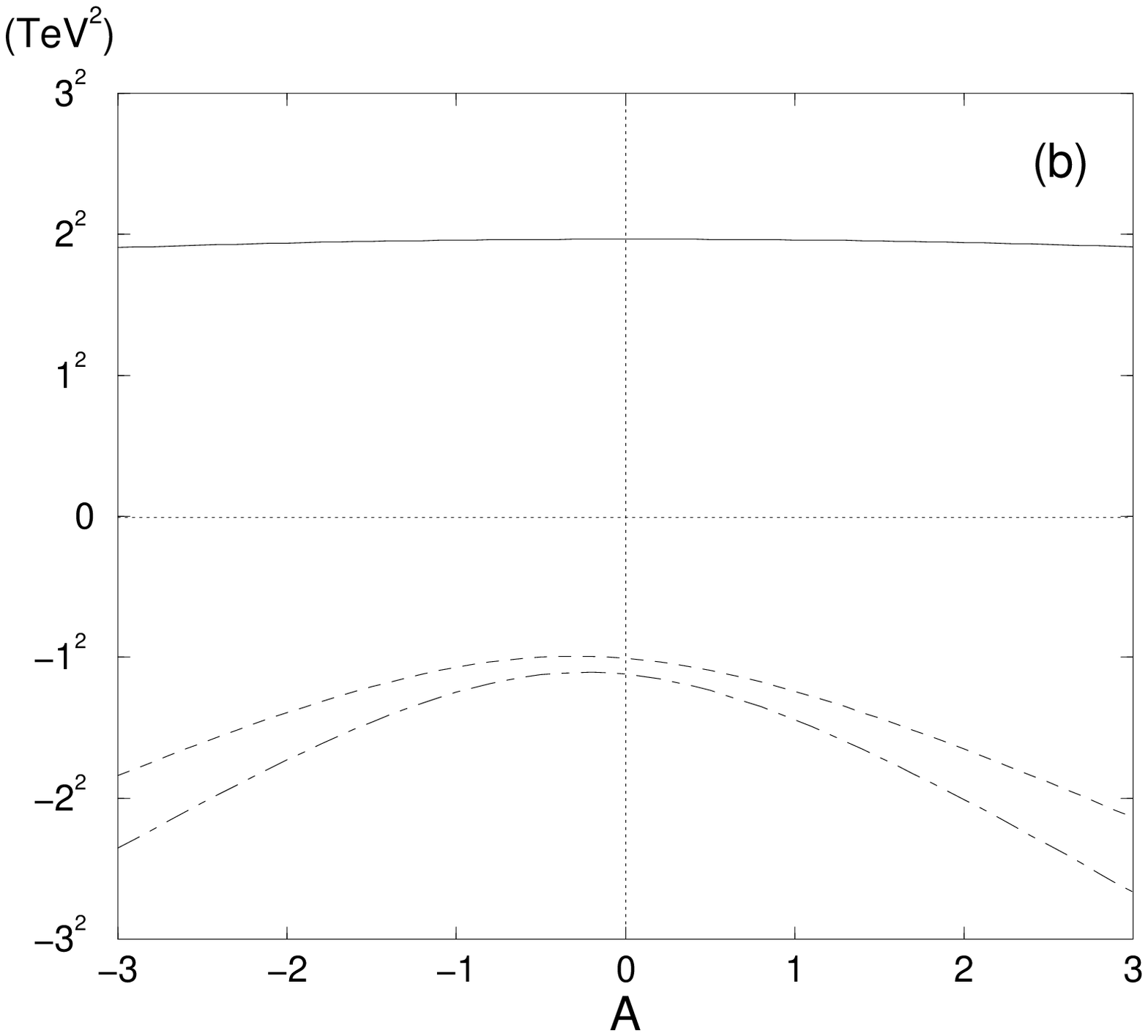,height=10cm}


\leavevmode 
\psfig{file=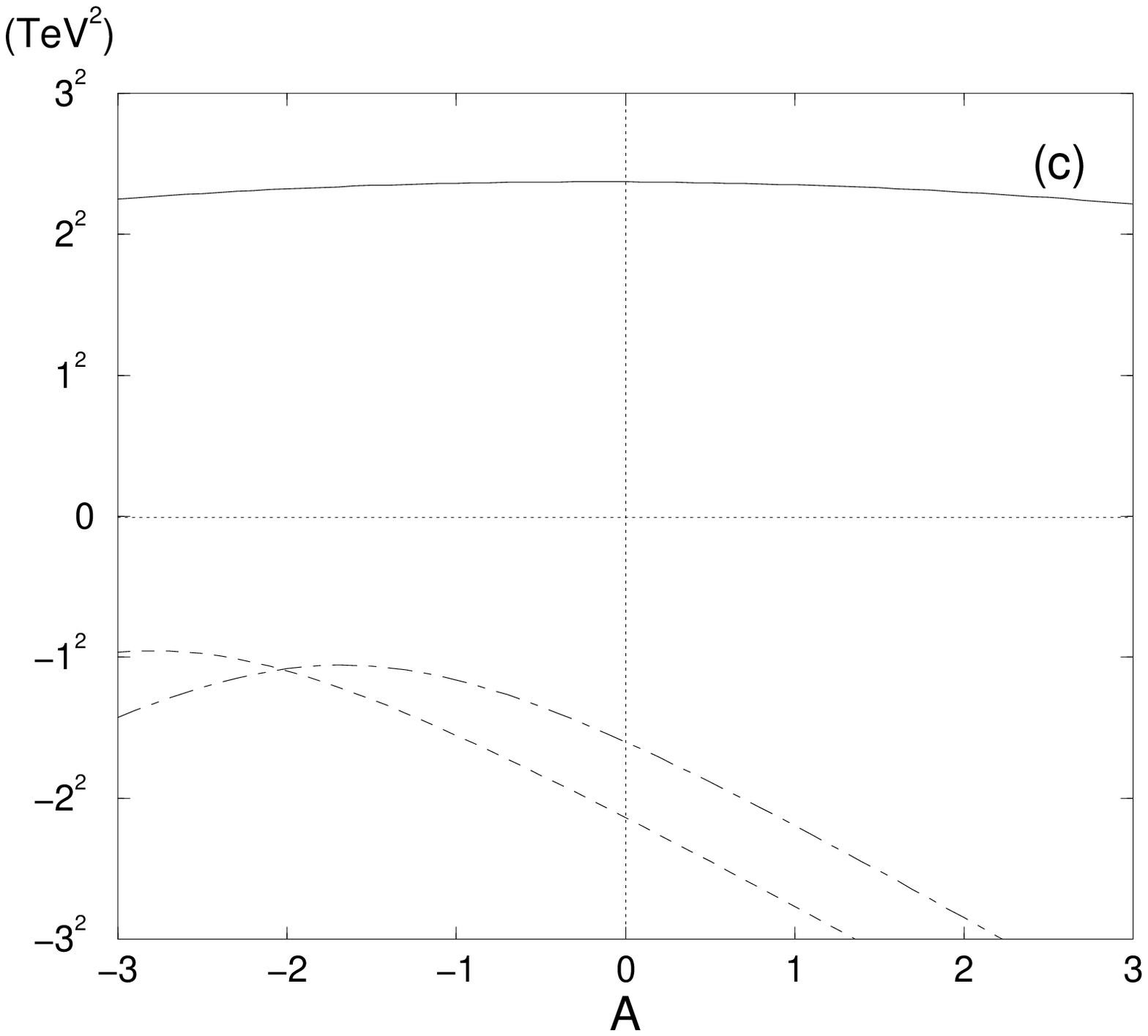,height=10cm}
\end{center}

\end{figure}

     The mass-squared parameters $\MHs1$, $\MHs2$, and $\MSs$ 
are shown, as functions of $A$, for the three parameter sets (a), 
(b), and (c) of Table \ref{parameters}   
in Figs. \ref{rgemass}(a), \ref{rgemass}(b), and \ref{rgemass}(c), 
respectively, with $\ln^X=\lh^X=0.2$.    
The gravitino mass is fixed as $\mg=2$ TeV.  
At the low energy scale, receiving large 
quantum corrections, $\MHs2$ and $\MSs$ could 
become much smaller than the universal value $\mg^2$.    
These corrections strongly depend on $\eta_u^X$, $\lk^X$, $\tilde m$, 
and $A$.  
In Fig. \ref{rgemass}(a), $\MHs2$ and $\MSs$ become negative only 
for $|A|\gsim 2$.  
However, as $\eta_u^X$ and $\lk^X$ increase, $\MHs2$ and $\MSs$ 
decrease, respectively.  
These masses-squared also become small for larger values of $\tilde m$.   
In Figs. \ref{rgemass}(b) and \ref{rgemass}(c), $\MHs2$ and $\MSs$ 
are negative for any value of $A$.  
If $\lk^X$ is larger than $\eta_u^X$, an inequality 
$\MSs<\MHs2$ holds at the low energy scale.  
On the other hand, $\MHs1$ is not much different from $\mg^2$.  

     We see from Figs. \ref{yukawa}, \ref{trilinear}, 
and \ref{rgemass} that the gauge symmetry of the vacuum at high 
energy scales can be spontaneously broken at low energy scales 
through radiative corrections.   
Furthermore, certain parameter values at a high energy scale, 
with the masses-squared and the trilinear coupling constants of 
scalar fields being universal, lead to   
the values of $\MHs1$, $\MHs2$, $\MSs$, $\lh$, and $B_H$   
which give a plausible vacuum around the electroweak energy scale.  
As explicit examples, we show in Table \ref{rgetab} the parameter values 
at $M_X$ which give low-energy vacua consistent with experimental results.     
Since there are many input parameters, for simplicity, we have fixed   
the trilinear coupling constant and 
the Yukawa coupling constants as  
$A=-1$ and $\eta_u^X=\ln^X=\lk^X=0.2$.    
The values of $\mg$ and $\tilde m$ are set for $\mg=1$, 2 TeV 
and $\tilde m=0.3$, 0.5 TeV.  
The resultant values of $v_2/v_1$, $v_s$, $M_{Z_2}$, $R$, and 
the Higgs boson masses at $M$ are also given together 
with the masses of the $W$ boson $M_W$, the top quark $m_t$, 
and the bottom quark $m_b$.  
This model is compatible with the supersymmetry breaking mechanism 
based on $N=1$ supergravity.  

\section{Conclusions}
     
     The MSSM involves the problems on proton lifetime, 
neutrino masses, and the $\mu$ term.  
These problems may be solved by theories at 
very high energy scales, which however are mere conjectures  
and mostly untestable at present.  
On the contrary, the solutions of the problems may reside 
in theories around an energy scale of the MSSM or the SM.  
Then, a new model beyond the MSSM should exists.  
Such a model is rigidly constructed on solid ground of various 
experimental results.   
Its predictions can be examined at experiments in the near future.  

     To solve the problems of the MSSM without invoking uncertain 
theories, we have constructed a supersymmetric standard model 
based on SU(3)$\times$SU(2)$\times$U(1)$\times$U$'$(1) gauge 
symmetry and $N=1$ supergravity.  
In this model, the solutions are given consistently 
within the framework of the model.  
The interactions of dimension four or five which violate baryon 
number conservation are not allowed  by the gauge symmetry, 
leading to an adequately long lifetime of the proton.  
The gauge symmetry also prescribes 
the existence of right-handed neutrinos and an SU(2)-singlet 
Higgs boson whose \vev\ $v_s$ induces breaking of 
the U$'$(1) symmetry.  
The neutrino masses receive contributions of Dirac type and 
those of Majorana type generated by the large \vev\ $v_s$.  
The ordinary neutrinos then have tiny masses.  
The $\mu$ term of the MSSM is replaced by a trilinear coupling of the 
Higgs superfields, and the effective $\mu$ parameter is given by $v_s$.   
A typical mass scale of this model is of order 1 TeV.  
Some fine-tuning of the parameters is necessary for correct 
breaking of the electroweak gauge symmetry.  
On the other hand,  the EDMs of the neutron and the electron 
are predicted within their experimental bounds 
without fine-tuning much $CP$-violating phases.  

     The constructed model gives predictions different from the MSSM 
in various phenomenological aspects.  
An extra neutral gauge boson couples to all the quarks and leptons.  
There exists a stable fermion which is nontrivially transformed 
under SU(3) and U$_{EM}$(1).  
Lepton number is not conserved in the interactions of the neutrinos 
or the scalar neutrinos.  
Some scalar particles mediate flavor-changing neutral current 
at the tree level.   
The experimental examination of these predictions could be performed 
in the near future.  

     Implications of this model for theories at higher  
energy scales have also been studied.  
The gauge symmetry breaking is induced by radiative 
corrections.  
The masses-squared and the trilinear coupling constants 
of scalar fields could be universal at an energy 
scale not much smaller than the Planck mass, which is 
consistent with the mechanism of supersymmetry breaking 
based on $N=1$ supergravity.  
The gauge coupling constants are not unified below the Planck mass 
scale, unless the particle contents are modified.

\acknowledgements
\smallskip

     We thank T. Watanabe for his advice on computer calculations.  
One of us (M.A.) acknowledges the Japan 
Society for the Promotion of Science for financial support.   
The work of M.A. is supported in part by 
the Grant-in-Aid for Scientific 
Research from the Ministry of Education, Science and Culture, Japan.  

\appendix
\section*{A}

\def\nn{\nonumber}
\def\dmu{\mu\frac{d}{d\mu}}

     The RGEs are listed below. The gauge and Yukawa coupling 
constants are expressed by $\alpha\equiv g^2/4\pi$ and  
$E\equiv\eta^2/4\pi$, $L\equiv\lambda^2/4\pi$.  
 
{\it The gauge coupling constants and the gauge fermion masses:}   
\begin{eqnarray}
\dmu\alpha_a &=& \frac{b_a}{2\pi}\alpha_a^2,  \\
\dmu\tilde m_a &=& \frac{b_a}{2\pi}\alpha_a\tilde m_a,  
\end{eqnarray}
where $b_3=0$, $b_2=3$, $b_1=15$, and $b_1'=15$ for 
SU(3), SU(2), U(1), and U$'$(1), respectively.  

{\it The masses-squared of the scalar fields:}  
\begin{eqnarray}
\dmu M_Q^2 &=& -\frac{2}{\pi}\left(\frac{4}{3}\a3\m3^2
+\frac{3}{4}\a2\m2^2 +\frac{1}{36}\a1\m1^2+\frac{1}{144}\ap\mp^2\right) 
 +\frac{1}{2\pi}\left(\frac{1}{6}\a1\xi+\frac{1}{12}\ap\xi'\right) 
\nn \\
&& +\frac{1}{2\pi}E_u\left(|A_u|^2\mg^2+\MHs2+M_Q^2+M_{U^c}^2\right) 
\nn \\
&& +\frac{1}{2\pi}E_d\left(|A_d|^2\mg^2+\MHs1+M_Q^2+M_{D^c}^2\right),  
\\
\dmu M_{U^c}^2 &=& -\frac{2}{\pi}\left(\frac{4}{3}\a3\m3^2
       +\frac{4}{9}\a1\m1^2 +\frac{1}{144}\ap\mp^2\right)  
+\frac{1}{2\pi}\left(-\frac{2}{3}\a1\xi+\frac{1}{12}\ap\xi'\right)  
\nn \\
&& +\frac{1}{\pi}E_u\left(|A_u|^2\mg^2+\MHs2+M_Q^2+M_{U^c}^2\right),  
\\
\dmu M_{D^c}^2 &=& -\frac{2}{\pi}\left(\frac{4}{3}\a3\m3^2
         +\frac{1}{9}\a1\m1^2+\frac{49}{144}\ap\mp^2\right)  
 +\frac{1}{2\pi}\left(\frac{1}{3}\a1\xi+\frac{7}{12}\ap\xi'\right)   
\nn \\
&& +\frac{1}{\pi}E_d\left(|A_d|^2\mg^2+\MHs1+M_Q^2+M_{D^c}^2\right),  
\\
\dmu M_L^2 &=& -\frac{2}{\pi}\left(\frac{3}{4}\a2\m2^2
           +\frac{1}{4}\a1\m1^2+\frac{49}{144}\ap\mp^2\right)  
 +\frac{1}{2\pi}\left(-\frac{1}{2}\a1\xi+\frac{7}{12}\ap\xi'\right) 
\nn \\
&& +\frac{1}{2\pi}E_{\nu}\left(|A_{\nu}|^2\mg^2+\MHs2+M_L^2+M_{N^c}^2\right) 
\nn \\
&& +\frac{1}{2\pi}E_e\left(|A_e|^2\mg^2+\MHs1+M_L^2+M_{E^c}^2\right),  
\\
\dmu M_{N^c}^2 &=& -\frac{2}{\pi}\left(\frac{25}{144}\ap\mp^2 \right) 
 +\frac{1}{2\pi}\left(-\frac{5}{12}\ap\xi'\right)    
\nn \\
&& +\frac{1}{\pi}E_{\nu}\left(|A_{\nu}|^2\mg^2+\MHs2+M_L^2+M_{N^c}^2\right) 
\nn  \\
&& +\frac{2}{\pi}L_N\left(|B_N|^2\mg^2+\MSs+2M_{N^c}^2\right),  
\\
\dmu M_{E^c}^2 &=& -\frac{2}{\pi}\left(\a1\m1^2+\frac{1}{144}\ap\mp^2\right) 
 +\frac{1}{2\pi}\left(\a1\xi+\frac{1}{12}\ap\xi' \right)    
\nn \\
&& +\frac{1}{\pi}E_e\left(|A_e|^2\mg^2+\MHs1+M_L^2+M_{E^c}^2\right),  
\\
\dmu \MHs1 &=& -\frac{2}{\pi}\left(\frac{3}{4}\a2\m2^2
               +\frac{1}{4}\a1\m1^2+\frac{4}{9}\ap\mp^2\right) 
 +\frac{1}{2\pi}\left(-\frac{1}{2}\a1\xi-\frac{2}{3}\ap\xi'\right)  
\nn \\
&& +\frac{3}{2\pi}E_d\left(|A_d|^2\mg^2+\MHs1+M_Q^2+M_{D^c}^2\right) 
\nn \\
&& +\frac{1}{2\pi}E_e\left(|A_e|^2\mg^2+\MHs1+M_L^2+M_{E^c}^2\right) 
\nn \\
&& +\frac{1}{2\pi}L_H\left(|B_H|^2\mg^2+\MSs+\MHs1+\MHs2\right),  
\\
\dmu \MHs2 &=& -\frac{2}{\pi}\left(\frac{3}{4}\a2\m2^2
               +\frac{1}{4}\a1\m1^2+\frac{1}{36}\ap\mp^2\right) 
 +\frac{1}{2\pi}\left(\frac{1}{2}\a1\xi-\frac{1}{6}\ap\xi'\right) 
\nn \\
&& +\frac{3}{2\pi}E_u\left(|A_u|^2\mg^2+\MHs2+M_Q^2+M_{U^c}^2\right) 
\nn \\
&& +\frac{1}{2\pi}E_{\nu}\left(|A_{\nu}|^2\mg^2+\MHs2+M_L^2+M_{N^c}^2\right)  
\nn \\
&& +\frac{1}{2\pi}L_H\left(|B_H|^2\mg^2+\MSs+\MHs1+\MHs2\right),   
\\
\dmu \MSs &=& -\frac{2}{\pi}\left(\frac{25}{36}\ap\mp^2\right) 
 +\frac{1}{2\pi}\left(\frac{5}{6}\ap\xi'\right)  
\nn \\
&& +\frac{1}{\pi}L_N\left(|B_N|^2\mg^2+\MSs+2M_{N^c}^2\right) 
\nn \\
&& +\frac{1}{\pi}L_H\left(|B_H|^2\mg^2+\MSs+\MHs1+\MHs2\right) 
\nn \\
&& +\frac{3}{2\pi}L_K\left(|B_K|^2\mg^2+\MSs+M_K^2+M_{K^c}^2\right),  
\\
\dmu M_K^2 &=& -\frac{2}{\pi}\left(\frac{4}{3}\a3\m3^2
           +\frac{1}{9}\a1\m1^2+\frac{4}{9}\ap\mp^2\right)   
 +\frac{1}{2\pi}\left(\frac{1}{3}\a1\xi-\frac{2}{3}\ap\xi'\right)  
\nn \\
&& +\frac{1}{2\pi}L_K\left(|B_K|^2\mg^2+\MSs+M_K^2+M_{K^c}^2\right),  
\\
\dmu M_{K^c}^2 &=& -\frac{2}{\pi}\left(\frac{4}{3}\a3\m3^2
           +\frac{1}{9}\a1\m1^2+\frac{1}{36}\ap\mp^2\right)    
 +\frac{1}{2\pi}\left(-\frac{1}{3}\a1\xi-\frac{1}{6}\ap\xi'\right) 
\nn \\
&& +\frac{1}{2\pi}L_K\left(|B_K|^2\mg^2+\MSs+M_K^2+M_{K^c}^2\right),  
\end{eqnarray}
where $\xi=\sum Y_\phi M_\phi^2$ and $\xi'=\sum Q_\phi M_\phi^2$.  

{\it The Yukawa coupling constants:}  
\begin{eqnarray}
\dmu E_u &=& -\frac{2}{\pi}E_u\left\{\frac{4}{3}\a3+\frac{3}{4}\a2
+\frac{13}{36}\a1+\frac{1}{48}\ap-\frac{1}{4}(6E_u+E_d+E_{\nu}+L_H)\right\}, 
\\
\dmu E_d &=& -\frac{2}{\pi}E_d\left\{\frac{4}{3}\a3+\frac{3}{4}\a2
 +\frac{7}{36}\a1+\frac{19}{48}\ap-\frac{1}{4}(E_u+6E_d+E_e+L_H)\right\}, 
\\
\dmu E_\nu &=& -\frac{2}{\pi}E_\nu\left\{\frac{3}{4}\a2+\frac{1}{4}\a1
 +\frac{13}{48}\ap-\frac{1}{4}(3E_u+4E_{\nu}+E_e+4L_N+L_H)\right\}, 
\\
\dmu E_e &=& -\frac{2}{\pi}E_e\left\{\frac{3}{4}\a2
 +\frac{3}{4}\a1+\frac{19}{48}\ap-\frac{1}{4}(3E_d+E_{\nu}+4E_e+L_H)\right\}, 
\\
\dmu L_N &=& -\frac{2}{\pi}L_N\left\{\frac{25}{48}\ap
             -\frac{1}{4}(4E_{\nu}+10L_N+2L_H+3L_K)\right\}, 
\\
\dmu L_H &=& -\frac{2}{\pi}L_H\left\{\frac{3}{4}\a2
            +\frac{1}{4}\a1+\frac{7}{12}\ap \right. \nn \\
&& \left.-\frac{1}{4}(3E_u+3E_d+E_{\nu}+E_e+2L_N+4L_H+3L_K)\right\}, 
\\
\dmu L_K &=& -\frac{2}{\pi}L_K\left\{\frac{4}{3}\a3
  +\frac{1}{9}\a1+\frac{7}{12}\ap-\frac{1}{4}(2L_N+2L_H+5L_K)\right\}.  
\end{eqnarray}

{\it The trilinear coupling constants:} 
\begin{eqnarray}
\dmu A_u &=& -\frac{2}{\pi}\left(\frac{4}{3}\a3\m3+\frac{3}{4}\a2\m2
  +\frac{13}{36}\a1\m1+\frac{1}{48}\ap\mp\right)\frac{1}{\mg} 
\nn \\
&& +\frac{1}{2\pi}(6A_uE_u+A_dE_d+A_{\nu}E_{\nu}+B_HL_H), 
\\
\dmu A_d &=& -\frac{2}{\pi}\left(\frac{4}{3}\a3\m3+\frac{3}{4}\a2\m2
      +\frac{7}{36}\a1\m1+\frac{19}{48}\ap\mp\right)\frac{1}{\mg}  
\nn \\
&& +\frac{1}{2\pi}(A_uE_u+6A_dE_d+A_eE_e+B_HL_H), 
\\
\dmu A_\nu &=& -\frac{2}{\pi}\left(\frac{3}{4}\a2\m2+\frac{1}{4}\a1\m1
           +\frac{13}{48}\ap\mp\right)\frac{1}{\mg}  
\nn \\
&& +\frac{1}{2\pi}(3A_uE_u+4A_{\nu}E_{\nu}+A_eE_e+4B_NL_N+B_HL_H), 
\\
\dmu A_e &=& -\frac{2}{\pi}\left(\frac{3}{4}\a2\m2
          +\frac{3}{4}\a1\m1+\frac{19}{48}\ap\mp\right)\frac{1}{\mg}  
\nn \\
&& +\frac{1}{2\pi}(3A_dE_d+A_\nu E_\nu+4A_eE_e+B_HL_H), 
\\
\dmu B_N &=& -\frac{2}{\pi}\left(\frac{25}{48}\ap\mp\right)\frac{1}{\mg} 
\nn \\
&& +\frac{1}{2\pi}(4A_{\nu}E_{\nu}+10B_NL_N+2B_HL_H+3B_KL_K), 
\\
\dmu B_H &=& -\frac{2}{\pi}\left(\frac{3}{4}\a2\m2+\frac{1}{4}\a1\m1
       +\frac{7}{12}\ap\mp\right)\frac{1}{\mg}  
\nn \\
&& +\frac{1}{2\pi}(3A_uE_u+3A_dE_d+A_{\nu}E_{\nu}+A_eE_e  
                   +2B_NL_N+4B_HL_H+3B_KL_K), 
\\
\dmu B_K &=& -\frac{2}{\pi}\left(\frac{4}{3}\a3\m3
      +\frac{1}{9}\a1\m1+\frac{7}{12}\ap\mp\right)\frac{1}{\mg}  
\nn \\
&& +\frac{1}{2\pi}(2B_NL_N+2B_HL_H+5B_KL_K).   
\end{eqnarray}

%

\newpage
%
%
%
%
%
%
%

\pagebreak

\pagebreak

\pagebreak

\pagebreak 

%
%
%
\begin{table}
\caption{Particle contents and their quantum numbers.  
$i=1,2,3$; $j=1,..,n_j$; $k=1,..,n_k$; $l=1,..,n_l$.  
        }
\label{particles}
\begin{tabular}{l l c c l}
   &  SU(3) & SU(2) & U(1) & U$'$(1) \\  
\hline
$Q^i$    & 3   & 2 & $\frac{1}{6}$  & $\QQ$  \\ 
$U^{ci}$    & $3^*$ & 1 & $-\frac{2}{3}$ & $\QU$  \\ 
$D^{ci}$    & $3^*$ & 1 & $\frac{1}{3}$  & $\QD$  \\     
$L^i$    & 1   & 2 & $-\frac{1}{2}$ & $\QL$  \\ 
$N^{ci}$    & 1   & 1 &      0         & $\QN$ \\
$E^{ci}$    & 1   & 1 &      1         & $\QE$  \\
$H_1^j$  & 1   & 2 & $-\frac{1}{2}$ & $\QH1$  \\
$H_2^j$  & 1   & 2 & $\frac{1}{2}$  & $\QH2$  \\
$S^k$    & 1   & 1 &      0         & $\QS$   \\
$K^l$    & 3   & 1 & $Y_K$  & $\QK$  \\
$K^{cl}$ & $3^*$ & 1 & $-Y_K$ & $\QKc$  \\ 
\end{tabular}
\end{table}
\begin{table}
\caption{$\Up$ charges of the superfields.  $i=1,2,3$.}
\label{charges}
\begin{tabular}{c c c c c c}
$Q^i$ & $U^{ci}$ & $D^{ci}$ & $L^i$ & $N^{ci}$ & $E^{ci}$      \\ 
\hline
$\frac{1}{12}$ & $\frac{1}{12}$ & $\frac{7}{12}$ & $\frac{7}{12}$ & 
$-\frac{5}{12}$ & $\frac{1}{12}$ \\
\hline
$H_1^i$ & $H_2^i$ & $S^i$ & $K^i$ & $K^{ci}$ &       \\ 
\hline
$-\frac{2}{3}$ & $-\frac{1}{6}$ & $\frac{5}{6}$ & $-\frac{2}{3}$ & 
$-\frac{1}{6}$ &  
\end{tabular}
\end{table}
\begin{table}
\caption{The parameter values of the potential and the outcomes.  
$|B_H\lh\mg|=0.1$ TeV.}
\label{masses}
\begin{tabular}{c c c c c}
                   &  (i)        &  (ii)       &  (iii)      &  (iv)  \\
\hline
$\lh$              &  0.1        &  0.1        &  0.3        &  0.3   \\
$\MHs1$ (TeV$^2$)  & $1.0^2$     & $2.0^2$     & $1.0^2$     & $2.0^2$  \\ 
$\MHs2$ (TeV$^2$)  & $0.20^2$    & $0.30^2$    & $-0.40^2$   & $-0.60^2$  \\ 
$\MSs$ (TeV$^2$)   & $-0.448^2$  & $-0.888^2$  & $-0.470^2$  & $-0.671^2$  \\   
\hline
$v_2/v_1$          & 5.7           & 11.5        & 6.7   &  18.2   \\ 
$v_s$ (TeV)        & 2.1           & 4.2         & 2.2   &  3.2  \\
$M_{Z_2}$ (TeV)    & 0.64          & 1.26        & 0.66  &  0.95  \\
$R$ & $3.9\times 10^{-4}$ & $1.3\times 10^{-4}$  
                          & $3.9\times 10^{-4}$  &  $2.4\times 10^{-4}$  \\
$M_{H^0}$ (TeV)    & 0.086         & 0.091       & 0.074  &  0.072  \\
                   & 0.63          & 1.26        & 0.66   &  0.95  \\
                   & 0.94          & 1.87        & 1.04   &  2.03  \\
$M_{A^0}$ (TeV)    & 0.94          & 1.87        & 1.04   &  2.03  \\ 
$M_{H^\pm}$ (TeV)  & 0.94          & 1.87        & 1.04   &  2.03  \\ 
\end{tabular}
\end{table}
\begin{table}
\caption{The values of $\eta_u^X$, $\lk^X$, and $\tilde m/\mg$ for  
numerical evaluations.}
\label{parameters}
\begin{tabular}{c c c c}
   &  (a) & (b) & (c)  \\  
\hline
$\eta_u^X$, $\lk^X$    & 0.1   & 0.3  & 0.1   \\ 
$\tilde m/\mg$         & 0.1   & 0.1  & 1     \\ 
\end{tabular}
\end{table}
\begin{table}
\caption{The parameter values at the high energy scale and 
the low energy outcomes. 
$A=-1$, $\eta_u^X=\ln^X=\lk^X=0.2$.}
\label{rgetab}
\begin{tabular}{c c c c c}
                   &  (i)        &  (ii)       &  (iii)      &  (iv)  \\
\hline
$\mg$ (TeV)       &    1.0     &  1.0    &   2.0    &  2.0    \\
$\tilde m$ (TeV)  &    0.3     &  0.5    &   0.3    &  0.5    \\ 
$\eta_d^X$        &    0.007   &  0.002  &   0.010  &  0.005  \\   
$\lh^X$           &    0.356   &  0.412  &   0.307  &  0.339  \\ 
\hline
$v_2/v_1$          & 3.2        & 2.1      & 5.2   &  3.7   \\ 
$v_s$ (TeV)        & 2.4        & 3.5      & 3.8   &  4.4   \\
$M_{Z_2}$ (TeV)    & 0.73       & 1.05     & 1.14  &  1.32  \\
$R$ & $1.2\times 10^{-4}$ & $2.1\times 10^{-6}$  
                          & $9.7\times 10^{-5}$  &  $5.3\times 10^{-5}$   \\
$M_{H^0}$ (TeV)    & 0.062     & 0.042    & 0.068   &  0.066  \\
                   & 0.73      & 1.05     & 1.15    &  1.33   \\
                   & 1.06      & 1.32     & 1.95    &  2.05   \\
$M_{A^0}$ (TeV)    & 1.06      & 1.32     & 1.95    &  2.05   \\ 
$M_{H^\pm}$ (TeV)  & 1.06      & 1.32     & 1.95    &  2.05   \\ 
$M_W$ (GeV)        & 80        & 82       &  74    &  79  \\ 
$m_t$ (GeV)        & 166       & 161     &  160   & 166  \\   
$m_b$ (GeV)        & 2.9       & 1.2      &  2.4   &  1.8  \\ 
\end{tabular}
\end{table}

\end{document}